\documentclass[%
 reprint,
 amsmath,amssymb,
 aps,
]{revtex4-2}

\usepackage{chemformula} 
\usepackage[T1]{fontenc} 

\begin{document}
\preprint{APS/123-QED}

\title{Penalty and auxiliary wave function methods for electronic Excitation in neural network variational Monte Carlo}

\author{Zixiang Lu}
\email{luzixiang@pku.edu.cn}
\affiliation{School of Physics, Peking University, Beijing 100871, P. R. China}

\author{Weizhong Fu}
\affiliation{School of Physics, Peking University, Beijing 100871, P. R. China }

\date{\today}

\begin{abstract}
This study explores the application of neural network variational Monte Carlo (NN-VMC) for the computation of low-lying excited states in molecular systems. Our focus lies on the implementation and evaluation of two distinct methodologies – the penalty method and a novel modification of the auxiliary wave function (AW) method – within the framework of the FermiNet-based NN-VMC package. Importantly, this specific application has not been previously reported.Our investigation advocates for the efficacy of the modified AW method, emphasizing its superior robustness when compared to the penalty method. This methodological advancement introduces a valuable tool for the scientific community, offering a distinctive approach to target low-lying excited states. We anticipate that the modified AW method will garner interest within the research community, serving as a complementary and robust alternative to existing techniques. Moreover, this contribution enriches the ongoing development of various neural network ansatz, further expanding the toolkit available for the accurate exploration of excited states in molecular systems.
\end{abstract}
\maketitle
\section{INTRODUCTION}
In the realms of chemistry and physics, the modeling of excited states stands as a crucial endeavor, essential for deciphering experimental measurements, such as those encountered in photospectroscopy. However, the prediction of excited states proves to be a formidable challenge compared to their ground state counterparts. Various computational methodologies have been developed to address this challenge, with time-dependent density-functional theory~\cite{doi:10.1142/2914, 10.1063/1.1904586, PhysRevLett.52.997} (TDDFT) emerging as a widely embraced approach. Additionally, methods such as multireference configuration interaction~\cite{Szalay2012,Lischka2018} (MRCI), coupled-cluster~\cite{EMRICH1981397,10.1063/1.464746,doi:10.1146/annurev.physchem.59.032607.093602} (CC), and algebraic diagrammatic construction to the second order~\cite{https://doi.org/10.1002/wcms.1206} (ADC(2)) are frequently employed, although each method, despite its efficacy, comes with inherent limiting factors.

Quantum Monte Carlo~\cite{RevModPhys.73.33,TOULOUSE2016285,CASINO} (QMC) techniques, notably variational and diffusion Monte Carlo (VMC and DMC), offer a potent alternative.The accuracy of these techniques can be heightened through the refinement of trial wave function quality. Notably, real-space Monte Carlo methods exhibit a relatively low scaling of $O(N_e^{3-4})$, where $N_e$ denotes the number of electrons, making them computationally efficient compared to other high-order deterministic wave function methods. While VMC and DMC have predominantly been utilized for ground state calculations, recent reports indicate successful extensions to excited states~\cite{Filippi2009,Dash2019,Send2011,doi:10.1021/acs.jctc.2c00769,doi:10.1021/acs.jctc.6b00508,10.1063/5.0024572,doi:10.1021/acs.jctc.8b00879,PhysRevLett.121.167204,10.1063/5.0030949,Entwistle2023}.

A significant stride in the domain of VMC and DMC is marked by the incorporation of neural network wave function ansatz, profoundly enhancing the accuracy of these methods~\cite{Hermann2023,Hermann2020, Ren2023, PhysRevResearch.2.033429, gerard2022goldstandard, scherbela2021solving, Li2022, PhysRevResearch.4.013021}. These neural network models represent many-body wave functions, and VMC is employed to train these networks in an unsupervised manner, adhering to the variational principle of quantum mechanics. Although initial efforts primarily concentrated on ground state problems, recent breakthroughs have extended the scope to include excited states~\cite{Entwistle2023, pfau_natural_2023}.

Noteworthy contributions in this realm include Entwistle et al.~\cite{Entwistle2023}'s utilization of a penalty method to compute excited states using the PauliNet wave function ansatz, and Pfau et al.~\cite{pfau_natural_2023}'s proposal of a general extension of neural networks for computing multiple excited states. Remarkably, Pfau et al.'s method avoids the need for empirical penalty parameters and can directly leverage optimization strategies designed for ground states. It is pertinent to mention that before the development of NN-VMC for molecules, its application had already been explored in model Hamiltonians, with Choo et al.~\cite{PhysRevLett.121.167204} reporting excited state calculations. The method used by Choo et al.~\cite{PhysRevLett.121.167204}, though bearing some similarity to the penalty method, is also devoid of empirical parameters. In this paper, we distinguish this approach as the auxiliary wave function (AW) method, details of which will be expounded upon in the method section.

Given the diversity of neural networks employed in excited state methods, it remains unclear whether variations in results arise from differences in the excited state methodologies or from discrepancies in the choice of ansatz. To contribute to the ongoing discourse, we have implemented the penalty method with the FermiNet ansatz and introduced modifications to the AW method to better align it with the framework of real-space neural network wave functions. This preprint aims to provide a detailed discussion on the implementation and performance of both the penalty and modified methods within the context of FermiNet-based NN-VMC, offering valuable insights for the scientific community.


\section{THEORY AND METHODS}

\subsection{FermiNet Ansatz And Ground State Optimization} \label{groundstate}
Central to our investigation is the FermiNet Ansatz, a robust wave function formulation designed for many-electron systems. Initially introduced in Ref.~\citenum{PhysRevResearch.2.033429} and further refined in Ref.~\citenum{2011.07125}, this Ansatz leverages a neural network with a substantial number of parameters and nonlinear activation functions between layers. This architectural flexibility enables the accurate targeting of both ground state and excited state wave functions.

In aligning with conventional Variational Monte Carlo (VMC) methods to target the ground state wave function, FermiNet minimizes the energy expectation value of the Ansatz. The expression is defined as:
\begin{equation}
\begin{aligned}
E_{V}(\theta) = \frac{\int dX \psi_\theta^*(X) \hat{H} \psi_\theta(X)}{\int dX \psi_\theta^*(X)  \psi_\theta(X)} = \frac{\int dXp_\theta(X)E_L^\theta(X)}{\int dX p_\theta(X)}  \label{exp_energy}
\end{aligned}
\end{equation}
where $\{\theta\}$ represents the Ansatz parameters, $\hat{H}$ is the system's Hamiltonian, and $X$ signifies the state (positions and spins) of the system. The probability distribution is denoted as $p_\theta(X) = |\psi_\theta(X)|^2$, the estimated expectation energy $E_V$ is approximated as the average value of the local energy $E_L^\theta(X)$ on $X_k$:

\begin{equation}
E_{V} \approx \bar{E}_L = \frac{1}{M} \sum_{k = 1}^M E_L^{\theta}(X_k)  \label{energyexp}
\end{equation}
Gradients of the expectation energy are estimated through similar methods:
\begin{equation}
\nabla_\theta E_{V}(\theta)= [(E_L - [E_L]_{p_\theta})\nabla_\theta ln\psi]_{p_\theta} \label{energygradient}
\end{equation}
where '$[]_{p_\theta}$' denotes a Monte Carlo sampling using the distribution of $p_\theta$. This expression, involving Monte Carlo sampling using the distribution $p_\theta$, facilitates the optimization of parameters through gradient descent methods to attain the ground state wave function and energy. For a more comprehensive algorithm, we refer the readers to Ref.~\citenum{PhysRevResearch.2.033429,RevModPhys.73.33,TOULOUSE2016285,CASINO}
\subsection{The Penalty Method}
%
%
Within the VMC framework, efficient calculation and optimization of objective functions form a cornerstone of our methodology. Previous works have explored different objective functions and optimization strategies to target excited state wave functions, including the penalty method utilized in Ref.~\citenum{10.1063/5.0030949, doi:10.1021/acs.jctc.2c00769} for the traditional ansatz and in Ref.~\citenum{Entwistle2023} for the PauliNet ansatz.
The objective function $O[\psi_\theta]$ designed for minimizing excited state wave functions in Ref.\citenum{10.1063/5.0030949} is expressed as:
\begin{equation}
O[\psi_\theta] = E[\psi_\theta] + \sum_i\lambda_i|S_i^\theta|^2  \label{objective}
\end{equation}
here:
\begin{equation}
S_i^\theta = \frac{\langle\psi_\theta|\psi_i\rangle}{\sqrt{\langle\psi_\theta|\psi_\theta\rangle\langle\psi_i|\psi_i\rangle}}  \label{overlap}
\end{equation}
represents the pairwise overlap between $\psi_\theta$ and $\psi_i$, acting as the penalty term. To target the $N_{th}$ excited state wave function, the preceding $N-1$ eigenstate wave functions must be obtained. The overlaps with the target wave function are then calculated and added to the objective function as penalties. By ensuring that each self-defined parameter $\lambda_i$ exceeds the energy difference $E_{i+1}-E_i$, the $(i+1)th$ excited wave function can be effectively targeted through the minimization of the objective function.

%
%

%
%

Illustrating its effectiveness, we consider the case of targeting the first excited state. After determining the ground state $\psi_0$, the objective function for acquiring the first excited state is expressed as $O[\psi_\theta] = E[\psi_\theta] + \lambda_0|S_0^\theta|^2$. Decomposing $\psi_\theta$ into a linear combination of all eigenstates $\psi_i$, the objective function is reformulated as:
\begin{equation}
O[c_i] = \frac{\sum_i c_i^2E_i + \lambda_0 c_0^2}{\sum_i c_i^2} \ge \frac{\sum_i c_i^2E_i + (E_1 - E_0) c_0^2}{\sum_i c_i^2} \ge E_1\label{O}
\end{equation}
The equality is achieved if and only if $\psi_\theta = \psi_1$.
Similar considerations apply to the global minimum of $O[\psi_\theta] = E[\psi_\theta] + \sum_i^{N-1}\lambda_i|S_i^\theta|^2$, which gives $\psi_\theta = \psi_N$.
\subsection{The Auxiliary Wave Function Method}
A departure from the dependence on a free parameter is explored in Ref.\citenum{PhysRevLett.121.167204}, where Choo et al.~\cite{PhysRevLett.121.167204} employed a Gram–Schmidt orthogonalization process before each optimization step. The strategy involved optimizing the orthogonalized wave function towards lower energy to specifically target the first excited state:
\begin{equation}
\psi = \psi_\theta - \lambda \psi_0 \label{orthogonal}
\end{equation}
Here, $\psi_\theta$ is derived from an independent ansatz with its parameters $\{\theta\}$, termed an auxiliary wave function. $\psi_0$ represents the optimized ground state wave function, and $\lambda = \langle\psi_0|\psi_\theta\rangle/\langle\psi_0|\psi_0\rangle$ serves as the orthogonalization coefficient. The computation of $\lambda$ is facilitated through the Monte Carlo method from the distribution of 
$|\psi_0|^2$:

\begin{equation}
\lambda = \frac{\langle\psi_0|\psi_\theta\rangle}{\langle\psi_0|\psi_0\rangle} = \big{[}\frac{\psi_\theta}{\psi_0}\big{]}_{|\psi_0|^2}
\label{lambda calculation}
\end{equation}

However, in the context of FermiNet, the recomputation of $\lambda$ before each optimization step proves to be time-consuming and inconvenient. Consequently, we propose a modification to this algorithm for improved convenience. The energy expectation of the orthogonalized wave function($\psi$) is expressed as:
%
\begin{equation}
\begin{aligned}
E_{V}(\psi) &= \frac{\langle\psi_\theta-\lambda\psi_0|\hat{H}|\psi_\theta-\lambda\psi_0\rangle}{\langle\psi_\theta-\lambda\psi_0|\psi_\theta-\lambda\psi_0\rangle} \\&= \frac{\langle\psi_\theta|\hat{H}|\psi_\theta\rangle+\lambda^2\langle\psi_0|\hat{H}|\psi_0\rangle-2\lambda\langle\psi_\theta|\hat{H}|\psi_0\rangle}{\langle\psi_\theta|\psi_\theta\rangle+\langle\psi_0|\psi_0\rangle-2\lambda\langle\psi_\theta|\psi_0\rangle}
\label{new_exp}
\end{aligned}
\end{equation}
where $\psi_0$ is precise enough to be considered the true ground state, therefore we have:
\begin{equation}
\hat{H}|\psi_0\rangle = E_0|\psi_0\rangle
\label{energyfunc}
\end{equation}
Substituting Eq.\ref{energyfunc} and Eq.\ref{lambda calculation} into Eq.\ref{new_exp}, we arrive at:
\begin{equation}
E_{V}(\psi) = \frac{E_\theta - E_0|S_0^\theta|^2}{1 - |S_0^\theta|^2}
\label{new_exp2}
\end{equation}
where $E_\theta = \langle\psi_\theta|\hat{H}|\psi_\theta\rangle/\langle\psi_\theta|\psi_\theta\rangle$ is the expectation energy of the auxiliary wave function $\psi_\theta$, and $S_\theta = \langle\psi_\theta|\psi_0\rangle/\sqrt{\langle\psi_0|\psi_0\rangle\langle\psi_\theta|\psi_\theta\rangle}$ is the pairwise overlap of $\psi_\theta$ and $\psi_0$. By considering $E_{V}(\psi)$ as the objective function($O[\psi_\theta] = E_V({\psi})$), the first excited state can be targeted. This approach can be extended to target the $N_{th}$ excited states by minimizing:
\begin{equation}
O[\psi_\theta] = \frac{E_\theta - \sum_{i = 0}^{N-1}E_i|S_i^\theta|^2}{1 - \sum_{i = 0}^{N-1}|S_i^\theta|^2}
\label{higher state}
\end{equation}
The minimized value of $O[\psi_\theta]$ then corresponds to the total energy of the $N_{th}$ excited state, and the wave function $\psi_N$ can be obtained as:
\begin{equation}
\psi_N = \psi_\theta - \sum_{i = 0}^{N-1}\frac{\langle\psi_\theta|\psi_i\rangle}{\langle\psi_i|\psi_i\rangle}\psi_i
\label{higher state wave}
\end{equation}
It is noteworthy that the pairwise overlaps in Eq.\ref{higher state} are the overlaps between $\psi_\theta$ and lower-state wave functions obtained from Eq.\ref{higher state wave}, rather than the auxiliary wave functions.

A key observation is that, in contrast to the penalty method, the AW method does not hinge on the choice of the free parameter $\lambda$. This feature marks a significant advantage in terms of computational efficiency and method robustness.

\subsection{Overlap Estimation}
In both methods, the objective functions crucially involve the energy expectation and pairwise overlap, denoted as $S_i^\theta$. The accurate computation of this pairwise overlap is paramount for the optimization of excited states. Following the approach of Entwistle et al.~\cite{Entwistle2023}, the pairwise overlap is determined by:
\begin{equation}
S_i^\theta = sgn(\psi_i)\times sgn(\psi_\theta)\times \sqrt{[\frac{\psi_i}{\psi_\theta}]_{|\psi_\theta|^2}[\frac{\psi_\theta}{\psi_i}]_{|\psi_i|^2}}
\label{overlap2}
\end{equation}


The derivative of $S_i^\theta$ with respect to the parameters $\{\theta\}$ can be expressed as:
\begin{equation}
\partial S_i^\theta = \frac{1}{S_i^\theta}[\frac{\psi_\theta}{\psi_i} - [\frac{\psi_i}{\psi_\theta}]_{|\psi_\theta|^2}\partial ln\psi_\theta ]_{|\psi_\theta|^2}\times [\frac{\psi_i}{\psi_\theta}]_{|\psi_\theta|^2}
\label{de_overlap2}
\end{equation}
where we have dropped the dependence of $\psi$ on the many-electron coordinate $X$ for clarity, `$[ ]_\rho$' implies that the coordinate $X$ is sampled from the normalized distribution $\frac{\rho(X)}{\int\rho(X)}$.
An alternative evaluation method by Pathak et al.~\cite{10.1063/5.0030949} was also explored, utilizing a different distribution $\rho = |\psi_\theta|^2 + |\psi_i|^2$ to compute the overlap. However, owing to the non-normalized nature of the wave function in the Ferminet Ansatz, this method necessitates an additional parameter for adjusting the distribution's weight, rendering it less convenient. Consequently, for the calculations underpinning this study, we exclusively relied on Eq.\ref{overlap2} and Eq.\ref{de_overlap2} to robustly estimate the overlap and its gradient. 
\section{RESULTS And Discussion}
\subsection{Application To Small Atoms And Molecules}
\begin{figure}[!htbp]
  \centering
  \includegraphics[width=8.0cm]{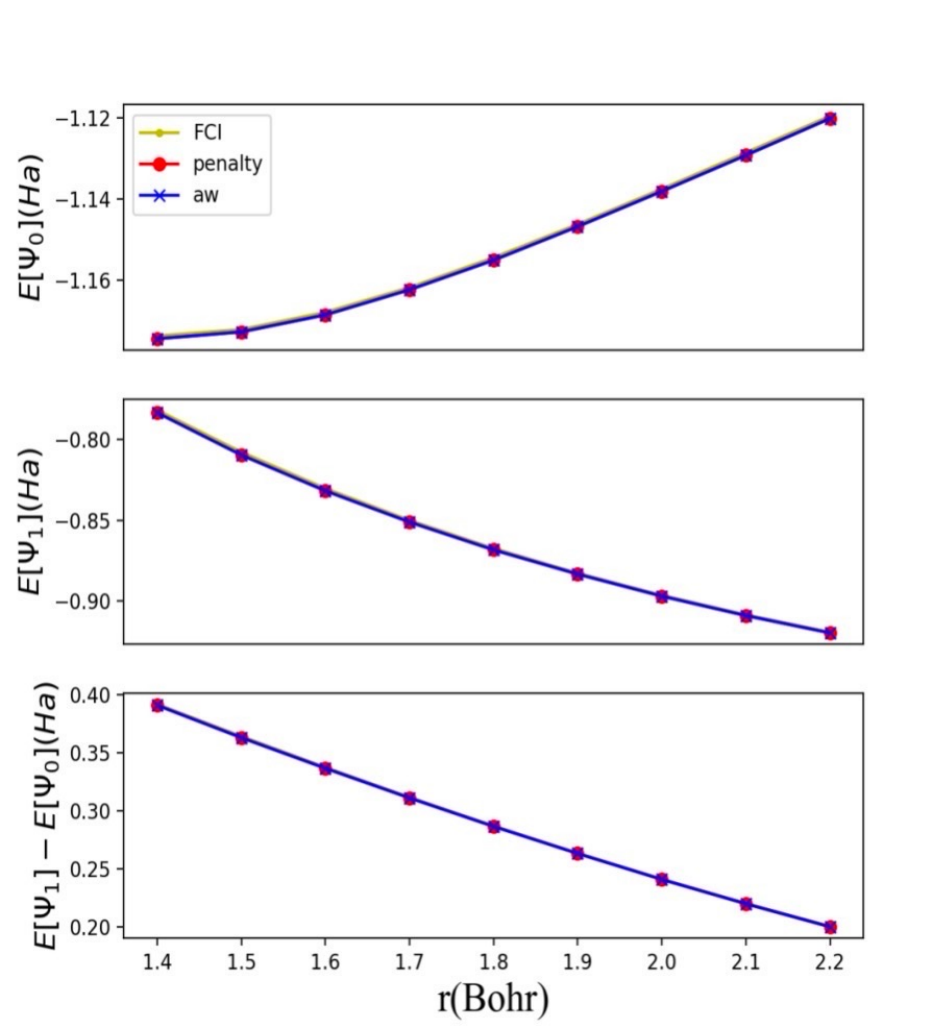}
  \caption{Total energy of the ground state and the first excited state for H2 with different bond lengths. We use the FCI result with the cc-pvqz basis set as the benchmark. Data of FCI results are taken from Ref.~\citenum{10.1063/5.0030949}
  }
  \label{H2}
\end{figure}

In our preliminary assessments, we subjected the penalty and auxiliary wave function (AW) methods to rigorous testing on the H2 molecule, a system with well-established benchmarks. Fig.\ref{H2} showcases the outcomes obtained for varying bond lengths of the H2 molecule, with the Full Configuration Interaction (FCI) results derived from the cc-pvqz basis set serving as our trusted benchmark.

For the penalty method, we meticulously tuned the free parameter, $\lambda$, to a value of 1. The comparative analysis reveals a commendable agreement between the penalty and AW results and the FCI benchmark. The maximum deviation observed in ground state energy stands at a mere 0.7 millihartree, showcasing a remarkable accuracy with an average deviation of 0.6 millihartree. Likewise, for the first excited state energy, we observe a deviation of 1.3 millihartree with an average deviation of 0.9 millihartree. The excitation energy shows a modest deviation of 0.9 millihartree, maintaining an average deviation of 0.4 millihartree. It is noteworthy that, given the diminutive nature of the H2 system, the ground state wave function can be precisely targeted, resulting in a notably small variance.
\begin{figure*}[!htbp]
  \centering
  \includegraphics[width=18.0cm]{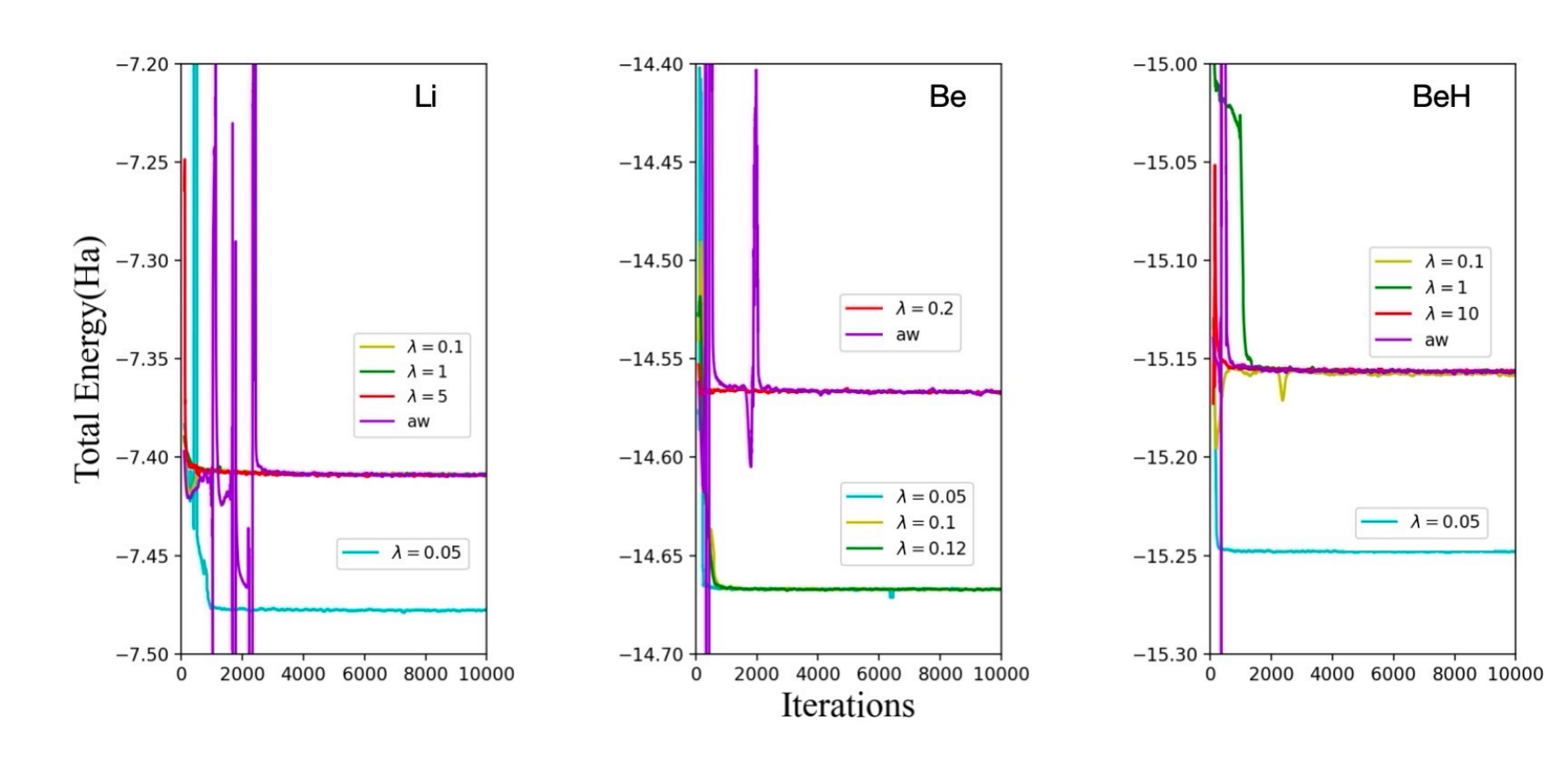}
  \caption{\textbf{Collapse of the optimization}. Optimization of the first excited state of Li, Be and BeH with different values of $\lambda$ are shown. Penalty optimization with unreasonable choices of $\lambda$ collapse to the ground state
  }
  \label{Li_Be_BeH}
\end{figure*}
To elucidate the impact of the empirical parameter in the penalty method, we conducted a comprehensive examination by employing different values of $\lambda$ and detailing their optimization procedures. For comparative purposes, we integrated the optimization results of the Auxiliary Wave (AW) method within the same figure. Our illustrative examples encompass Li, Be, and BeH, each known for its distinctive excitation dynamics.

The first excitations for these systems are as follows:
Li: 0.06795 Hartree (penalty method with $\lambda$ = 2 Hartree), 0.06790 Hartree (AW method)
Be: 0.1002 Hartree (penalty method with $\lambda$ = 2 Hartree), 0.1001 Hartree (AW method)
BeH: 0.0914 Hartree (penalty method with $\lambda$ = 2 Hartree), 0.0914 Hartree (AW method)
A critical observation emerges when using small values of $\lambda$ relative to the excitation energy, leading to optimization collapse to the ground state. Specifically, for Li, a $\lambda$ of 0.05 Hartree (smaller than the excitation energy) triggers collapse. For Be, $\lambda$ values of 0.05 and 0.1 Hartree (both smaller than the excitation energy) result in collapse, while a $\lambda$ of 0.12 Hartree (larger than the excitation energy) still collapses. Similarly, for BeH, a $\lambda$ of 0.05 Hartree (smaller than the excitation energy) induces collapse.

Entwistle et al.~\cite{Entwistle2023} introduced a modified penalty term to mitigate this issue, but its optimization remains contingent on the choice of the empirical parameter. In stark contrast, the modified AW method does not hinge on $\lambda$ and exhibits resilience against collapse to the ground state. In instances where the auxiliary wave function approaches ground state collapse, the numerator and denominator of Eq.\ref{new_exp2} both approach zero, causing the loss function to intermittently surge to extremely high values. This behavior, as depicted in Fig. \ref{Li_Be_BeH}, indicates substantial fluctuations throughout the optimization process. The robustness of the modified AW method, free from reliance on empirical parameters, presents a notable advantage over the penalty method in addressing optimization challenges.

\subsection{Time Consumption}
\begin{figure*}[!htbp]
  \centering
  \includegraphics[width=18.0cm]{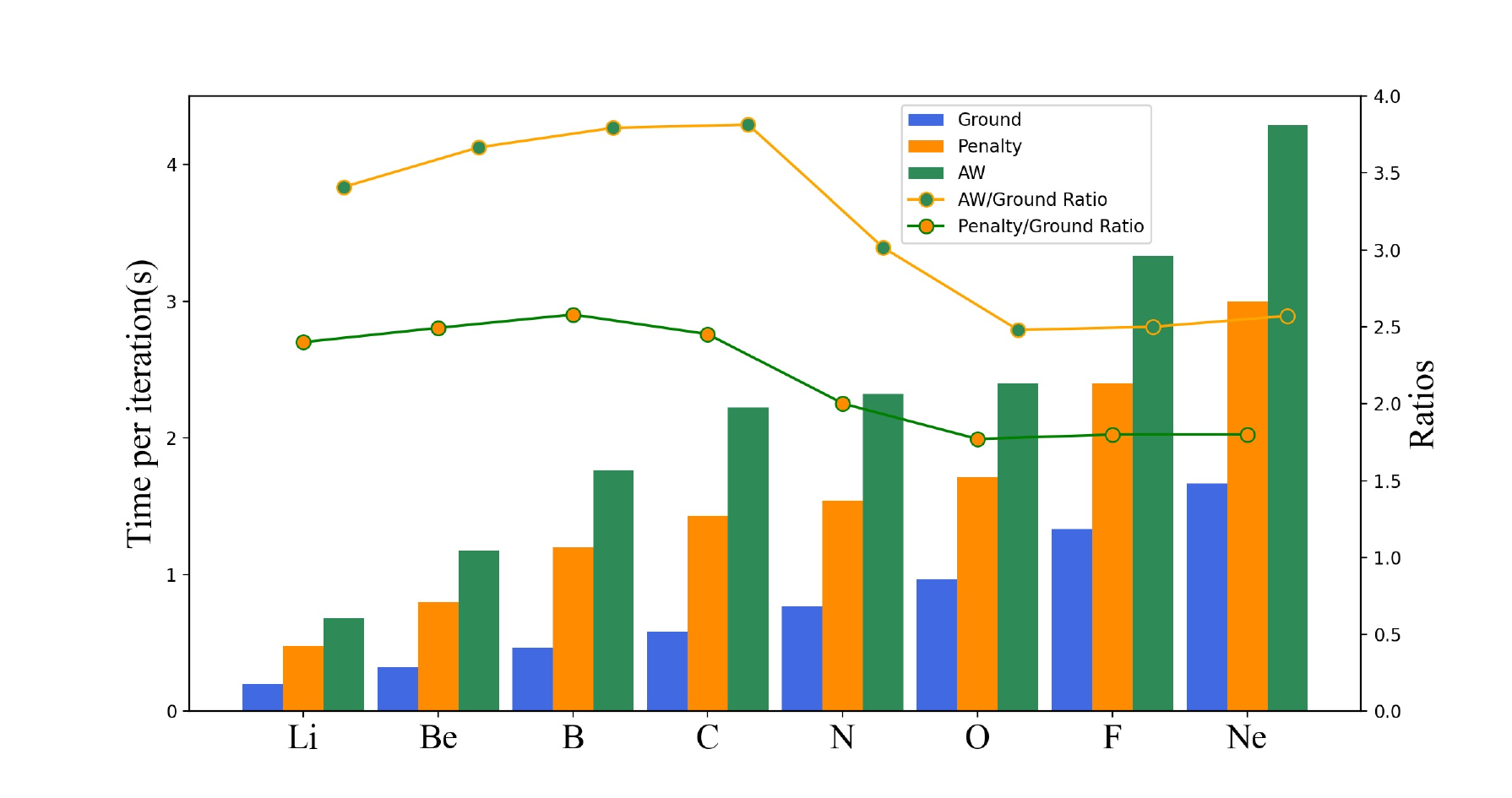}
  \caption{\textbf{Time consumption for atoms}.  Comparison of the runtime for one optimization iteration on atoms of the first row.
  }
  \label{time}
\end{figure*}
To project the time consumption for larger systems, we employed the same training configuration utilizing a single GPU and computed the time consumption per iteration for atoms of the first row. Each iteration encompassed 10 Markov Chain Monte Carlo (MCMC) steps, and the KFAC optimizer was employed. We calculated the time ratios for both the penalty and AW methods, denoted as $T_{\text{penalty}}/T_{\text{ground}}$ and $T_{\text{AW}}/T_{\text{ground}}$ respectively.
The observed time ratios for both methods exhibited stability. Notably, the AW method demonstrated a relatively higher time consumption per iteration due to the additional computational load imposed by the orthogonalization procedure. Empirically, it is recognized that the time consumption for ground state calculations can be accurately modeled by a cubic fit~\cite{PhysRevResearch.2.033429}. Consequently, we infer that the estimation of excited states also follows a time scale of approximately $O(N^2)$ for systems that do not exhibit significant size expansion.

\section{DISCUSSION}
In our exploration of NN-VMC utilizing the Ferminet wave function ansatz, we meticulously implemented both the penalty method and a modified version of the Auxiliary Wave (AW) method. These methods were applied to target the first few excited states of selected small systems, employing both the adam and kfac optimizers.

For small systems, we observed that both the penalty and the modified AW methods exhibit commendable accuracy in predicting excited states. However, the penalty method's stability is contingent on the choice of a suitable empirical parameter; selecting an inappropriate parameter may lead to instability. In contrast, the modified AW method demonstrates robustness, devoid of such sensitivity to empirical parameters.

The AW method showcases a distinct advantage in ensuring orthogonality among excited states—an attribute not guaranteed by the penalty method. However, a notable limitation of the AW method arises in its inability to provide the variance of the expectation energy during optimization. This restriction precludes the application of variance matching and variance extrapolation methods, limiting its utility in certain contexts.

Furthermore, employing the AW method necessitates optimizing the neural network ansatz to represent a linear combination of the ground state and the excited state. Conversely, the penalty method requires optimizing the ansatz to closely resemble the first excited state, a theoretically more challenging task.

In summary, while both methods demonstrate efficacy in predicting excited states for small systems, the modified AW method stands out for its stability across a range of empirical parameters and its ability ensure orthogonality. However, its limitations in direct acquirement of wave function should be considered in the context of specific research objectives. The penalty method, although potentially prone to instability, offers a distinct advantage in directly optimizing the ansatz to represent the excited states, albeit with the caveat of empirical parameter sensitivity. The choice between these methods should be made judiciously, considering the specific requirements and challenges posed by the system under investigation.

\begin{acknowledgements}
This work is supported by Ji Chen and ByteDance research. 
\end{acknowledgements}


\bibliography{main-demo}

\end{document}